\def\be{\begin{equation}}
\def\te{\end{equation}}
\def\bea{\begin{eqnarray}}
\def\nn{\nonumber}
\def\tea{\end{eqnarray}}

\documentstyle[12pt]{article}
\makeatletter 
\@addtoreset{equation}{section}
\makeatother  

\begin{document}
\title{Quantum Fluctuations, Decoherence of the Mean Field,
and Structure Formation in the Early Universe}
\author{E. Calzetta \\
{\small IAFE and FCEN, University of Buenos Aires, Argentina}\\
 B. L. Hu \\
{\small Department of Physics, University of Maryland, College Park,
MD 20742, USA}\\
{\small School of Natural Sciences, Institute for Advanced Study, Princeton,
NJ 08540, USA}}
\date{\small {\it (IASSNS-HEP-95/38, UMDPP 95-083, May 9, 1995)}}
\maketitle

\newpage

\section{Introduction and Summary}

In this paper we shall examine, starting from first principles, under what
circumstances the fluctuations of a quantum field transmute into classical,
stochastic fluctuations. To do so we shall analyze the relationship between
the phenomena of dissipation, fluctuation, noise and decoherence \cite
{Tsukuba}, first in an interacting scalar field theory in flat space time
\cite{Banff}, and then in the more complex but realistic case of a scalar
field interacting with gravitons in an expanding Universe.

The main motivation for this work is to develop the necessary tools to
analyze the quantum to classical transition of primordial density
fluctuations in the early universe. Indeed, this is the third of a series of
papers by the authors and collaborators on the quantum statistical theory of
structure formation. The first one \cite{Belgium} calls into question
conventional treatments of this issue, and focuses on decoherence and the
quantum origin of noise in stochastic inflation. The second one \cite{nfsg}
explains how noise and fluctuations originate from particle creation in
semiclassical gravity, and casts doubt on the conventional practise of
simplistically identifying quantum and classical fluctuations. In this paper
we will discuss how a
quantum mean field can be decohered by its own quantum fluctuations, turning
into a classical stochastic field. We will also explain how a proper
treatment of quantum and classical fluctuations can lead to a much improved
prediction of the density contrast in the inflationary cosmology.

\subsection{Outstanding Issues in the Quantum Theories of Galaxy Formation}

Let us begin by placing the present discussion within the larger framework
of theories of structure development in the universe. A standard mechanism
for galaxy formation is the amplification of primordial density fluctuations
by the evolutionary dynamics of spacetime \cite{Lifshitz,Bardeen}. In the
lowest order approximation the gravitational perturbations (scalar
perturbations for matter density and tensor perturbations for gravitational
waves) obey linear equations of motion. Their initial values and
distributions are stipulated, generally assumed to be a white noise
spectrum. In these theories, fashionable in the Sixties and Seventies, the
primordial fluctuations are classical in nature. The Standard model of
Friedmann-Lemaitre- Robertson-Walker (FLRW), where the scale factor of the
universe grows as a power of cosmic time, generates a density contrast which
turns out to be too small to account for the observed galaxy masses. The
observed nearly scale-invariant spectrum \cite{HarZel} also does not find
any easy explanation in this model \cite{Peebles,ZelNov}.

In the inflationary cosmology of the Eighties \cite{Guth,AlbSte,LindeInf} a
constant vacuum energy density of a quantum field $\Phi $, the inflaton,
drives the universe into a phase of exponential expansion, with the scale
factor $a(t)=a_0\exp (Ht)$, where $H=\dot a/a$ is the Hubble expansion
rate, assumed to be a constant for the de Sitter phase of the evolution
(a dot over a quantity stands for a derivative with respect
to cosmic time). The scalar field $\Phi $ evolves according to
the equation
\begin{equation}
\ddot \Phi +3H\dot \Phi +V^{\prime }[\Phi ]=0,
\end{equation}
where the potential $V[\Phi ]$ can take on a variety of forms, such as $\Phi
^4$ (Guth's original `old' inflation via tunneling \cite{Guth},
or Linde's chaotic inflation via rolling down the potential \cite{ChaoInf}),
Coleman-Weinberg (`new' inflation \cite{AlbSte,LindeInf}), or an
exponential form (for power law inflation \cite{LucMat}).
\footnote{As we shall discuss
in more detail below, agreement with the average amplitude of primordial
energy density fluctuations requires, in the conventional approaches, that
the scalar potential has a flat plateau, which generally is only possible if
the potential is fine-tuned for that purpose. For this reason, none of the
implementations of inflation proposed so far is regarded as totally
satisfactory.}
 Consider the `eternal inflation' stage where the universe has
locally a de Sitter geometry, with a constant Hubble radius (de Sitter
horizon) $l_h=H^{-1}$. The physical wavelength $l$ of a
mode of the inflaton field is $l=p^{-1}=a/k$, where $k$ is the wave number
of that mode. As the scale factor increases exponentially, the wavelengths
of many modes can grow larger than the horizon size. After the end of the de
Sitter phase, the universe begins to reheat and turns into a
radiation-dominated Friedmann universe with power law expansion $a(t)\sim
t^n $. In this phase, the Hubble radius grows much faster than the physical
wavelength, and some inflaton modes will reenter the horizon. The
fluctuations of these long-wavelength inflaton modes that go out of the de
Sitter horizon and later come back into the FLRW horizon play an important
role in determining the large scale density fluctuations of the early
universe, which in time seeded the galaxies \cite{GalForInf}.

With the exponential expansion in the de Sitter phase, any classical
primordial inhomogeneity will likely be redshifted out of existence by
the time the relevant modes leave the horizon, and one may wonder where such
fluctuations could arise. In the context of inflation,
Starobinsky \cite{StoInf} and others observed that the inflaton
field driving inflation is itself subject to quantum fluctuations,
which may provide the seeds for structure formation.

As a concrete example, in stochastic inflation \cite{StoInf}, the inflation
field is divided into two parts at every instant according to their physical
wavelengths, i.e.,

\begin{equation}
\label{split}\Phi (x)= \phi(x) + \psi(x).
\end{equation}
The first part $\phi $ (the `system field') consists of field modes whose
physical wavelengths are longer than the de Sitter horizon size $p< \epsilon H$
($\epsilon \approx 1$). The
second part $\psi $ (the `environment field') consists of field modes whose
physical wavelengths are shorter than the horizon size $p> \epsilon H$.
Inflation
continuously shifts more and more modes of the environment field into the
system, stretching their physical wavelengths beyond the de Sitter horizon
size. It is often stated that this process generates an effective
interaction between system and environment, in spite of the fact that
the fields in these models are free, leaving no chance for any mode -mode
coupling. The system field would then be randomly driven by
the unknown environment field, developing stochastic fluctuations which are
the required primordial fluctuations.

While this overall picture is generally agreeable, not least because of its
qualitative depictive power (it makes present day structures correspond to
near-Planckian scales early enough in the inflationary period, whereby the
physics of these fluctuations is expected to be mostly model independent),
it is important not to overlook its basic shortcomings, like the
oversimplified treatment of the quantum to classical transition and the
unneccesarily overweening role it ascribes to our own scientific interests
in defining the system - environment split. As we shall show below,
in taking these conceptual points seriously one can significantly improve
on the quantitative predictions on inflationary models. Let us now discuss
these issues in more detail, as an introduction to the main body of this
paper.

\subsubsection{How quantum fields acquire classical stochastic behavior:
Decoherence}

Consider Starobinsky's model \cite{StoInf} of a free, massless,
minimally-coupled inflaton field. Using the separation (\ref{split}), the
equation of motion for the system field $\phi $ is given by

\begin{equation}
\ddot \phi (t)+3H\dot \phi +V^{\prime }(\phi )=\xi (t)
\end{equation}
where $\xi $ is a white noise originating from the high frequency modes of
the bath field $\psi $ with properties,
\begin{equation}
<\xi (t)>=0,~~~~<\xi (t)\xi (t^{\prime })> \simeq \delta (t-t^{\prime })
\end{equation}
The common belief is that the short wavelength field modes (the bath)
contribute a white noise source to a classical Langevin equation governing
the long-wavelength (system) field modes. A Fokker-Planck equation can also
be derived which depicts the evolution of the probability distribution of
the scalar field $P(\phi ,t)$ \cite{Graziani}. Much recent effort is devoted
to the solution of this stochastic equation or its related Fokker-Planck
equation for descriptions of the inflationary transition and galaxy
formation problems. Although this scenario leads to the prediction of an
essentially scale- free distribution of density fluctuations, consistent
with the observational data \cite{HarZel}, and in spite of continued
efforts, no  satisfactory  implementation of these ideas has been
proposed so far.

Note that in transforming a quantum field theoretic problem to a classical
stochastic mechanics problem as in here, two basic assumptions are made:

1) The low frequency scalar field modes (the system field) behave
classically, and

2) The high frequency quantum field modes (the environment field) behave
like a white noise.

Most previous researchers seem to hold the view that the first condition is
pretty obvious \cite{GuthPi} and that the second condition can be easily
proven. One of us \cite{cgea} challenged this view and called attention to
the need for building a sounder foundation to the quantum theory of
structure formation. A rigorous program of investigation was outlined in
\cite{Belgium} with quantum open system concepts \cite{qos} and the
influence functional formalism \cite{if}. It was stressed that on the issue
of quantum to classical transition, one needs to consider the decoherence
process \cite{envdec,conhis}, and on the issue of noise, one needs to trace
its origin to the quantum field interactions and the coarse-graining
measures involved. These two issues are interrelated, as the noise in the
environment is what decoheres the system and endows it with a classical
stochastic dynamics.

Technically, the dynamics of the system is described by an influence action,
which is generally both complex and nonlocal (it becomes local for the
rather special case of an ohmic bath, but this is unimportant to our present
concerns). The imaginary part of this influence action is related to both
decoherence and spontaneous fluctuations in the unfolding of the system
variables; thus, decoherence is always associated with noisiness \cite
{GelHar2,HMLA}. The nonlocal part is associated with dissipation, and it is
related to the imaginary part through the fluctuation-dissipation relation
\cite{fdr,fdrgr,fdrsc}.
Thus, in a nonlinear theory,  decoherence, fluctuation
and dissipation are interrelated aspects
of the same phenomenon \cite{Tsukuba,GelHar2,nfsg}. We can visualize this as
dissipation representing the average action of the bath on the system, while
fluctuation describing the departures from the average. The nonlinear
interaction also creates correlations, whose severing upon tracing of the
bath degrees of freedom induces decoherence.
\footnote {In order to prevent misunderstanding,
let us observe that in the case a
quantum oscillator following an accelerated trajectory and coupled to a
quantum scalar field, where the influence action may be non trivial even if
the Hamiltonian is quadratic \cite{rindler}, the Hamiltonian is not diagonal
if expressed in terms of ``oscillator'' and ``field'' degrees of freedom;
actually, the transformation from these ``naive'' modes to those that
diagonalize the Hamiltonian is non analytic in the coupling strenght among
field and oscillator, so we can not even speak of a ` `weakly coupled''
regime. Therefore, in this case we may say that the non triviality of the
influence action is induced by an arbitrary partition of the degrees of
freedom in relevant and otherwise.}

In \cite{Belgium} a model of two interacting fields representing the system
and the bath is used to derive the (functional) Langevin equation and the
correlator of the (colored) noise. Further work need be carried out in
finding solutions to these stochastic equations for galaxy formation
considerations \cite{Yi}. A recent work along lines similar to ours
is that of Buryak \cite{Buryak}.

Given the complexity of the quantum to classical transition issue, one
may be tempted, as is indeed the case for most researchers on this topic,
to forget all about it, simply expand the quantum inflaton field in
any suitable set of modes, and identify the density profile
with the amplitude of those modes.
However, some careful thought will reveal this position to be untenable.
To begin with, extracting the physically observable
field variable out of the basic quantum one is not always trivial, both are
related through the renormalization process \cite{chrisDJ}. Besides,
while one can describe the quantum fluctuations in the inflaton
field as a coherent superposition of localized fluctuations, this does not
imply a physical inhomogeneity, because different fluctuations are not
mutually exclusive, and the quantum state is homogeneous \cite{Mat}.
Only when these
fluctuations become mutually exclusive, through the process of decoherence,
and some of them are realized, by the equivalent of some `measurement'
process, will it be proper to speak of inhomogeneity in the Universe.
In other words, a quantum field may be expanded in any set of basic modes (for
example, Minkowsky or Rindler modes in flat spacetime),
but only one preferred set may describe the observable (classical) density
fluctuations. Which mechanism gives that particular set its special
character is a physical question (not unlike which criterium picks out the
preferred pointer basis in environment-induced state reduction \cite{envdec}),
and should better be answered on the basis of the dynamics of the system
itself.

\subsubsection{Coarse-graining a noninteracting field cannot generate noise}

Most discussions on the origin of primordial fluctuations in the literature
are confined to a free scalar field propagating on a fixed geometrical
background. This cannot, as we argued in \cite{Belgium} using stochastic
inflation as an example, generate any noise, and without noise the system
cannot decohere and become classical. It also misses out all the interesting
phenomena associated with changes of correlations in the system due to
nonlinearities. (Even those who properly account for the mixing of matter
and gravitational degrees of freedom, say, by employing gauge invariant
rather than canonical variables \cite{Bardeen}, often stop at the linearized
level, thus missing the dynamical contribution to decoherence and the
evolution of correlations.) To compound the situation, most people would
agree that the initial quantum state of the field should be read out of a
Hartle-Hawking - like ``wave function of the Universe'' \cite{qc,decQC}, which
predicts lack of correlations among different modes. This leaves the
inflation practitioner with only two alternatives, namely, either
consider several free scalar fields and add a mixing matrix \cite{decInf},
so that the relevant degrees of freedom are not those that diagonalize the
Hamiltonian, or else consider a time-dependent system-bath
split, so that the correlations are carried by the modes themselves, as they
switch labels from ``bath mode'' to ``system mode'' or vice versa \cite
{StoInf,decInf}. It should be observed that the first alternative detracts
from the predictive power of the model, by introducing the elements of the
mixing matrix as so many new free parameters. In the second approach,
the whole issue of structure formation seems to hang on the special way one
labels the modes which define the system. This subjective element we find
rather uneasy. There is a third alternative,
which is to assume that decoherence never occurred, at least at long enough
wavelengths \cite{leonid}. This we regard as an evasive way out. We would
prefer
to see the decoherence of a system as a consequence of its own dynamics.

\subsubsection{Over-production of density contrast and the fine-tuning problem}

In addition to the problem of deriving a classical stochastic equation from
quantum field theory, there is also the outstanding problem of
over-amplification of density contrast and unnatural constraining of the
field parameters. Recall that one distinct advantage of inflation is that it
provides a natural explanation of the scale-invariant Harrison-Zeldovich
spectrum \cite{HarZel}. But the excess amplitude in the density contrast is
still an unresolved problem. The density contrast $\delta \rho /\rho $ can
be shown to be related to the fluctuations of the scalar field $\delta \Phi $
approximately by \cite{GalForInf}

\begin{equation}
{\frac{{\delta \rho }}\rho }\approx {\frac{{H\delta \Phi }}{{<\dot \Phi >}}}
\end{equation}
where $<~~>$ denotes averaging over some spatial range. In the conventional
treatment (where quantum fluctuations are treated in the same capacity as
classical fluctuations), for the density contrasts to be within $10^{-5}$
when the modes enter the horizon, the coupling constant in the Higgs field
(of, say, a $\lambda \phi ^4$ theory in the standard Grand Unified models)
has to be fine-tuned to an unnaturally small value ($\lambda \sim 10^{-12}$).\\


\noindent In summary, there are two sets of outstanding issues:

1) How the long-wavelength modes become classical, and the quantum
fluctuations develop into classical perturbations, and

2) how to get the correct order of magnitude for the density contrast
without asssuming an unnatural value for the field parameters.

We shall show in this paper that these two issues are related to each other:
decoherence of long-wavelength modes by short wavelength modes (as done in
\cite{Belgium}), or a mean field by its quantum fluctuations (done here),
gives rise to a classical stochastic evolution with noise properly
determined by the coupling between these two sectors. We shall consider a
more realistic model of gravitons coupled to the inflaton, and show that a
correct treatment of the relationship between quantum and classical
fluctuations
can provide a much improved estimate of the density contrast without
`fine-tuning'. We shall also investigate a different type of
system-environment split, namely, that between the mean field and its
fluctuations. This is in the spirit of the background field method used
frequently in quantum field theory. It is particularly relevant to
improvements beyond the mean field results in phase transition problems
\cite{sdqft}. We will also use this split as an example (the simplest) of the
correlation hierarchy discussed elsewhere in the context of decoherence
\cite{dch}.

It should be clear from the above that the proper identification of the
relevant system and its environment is an essential part of the analysis of
fluctuation generation in the early universe. This identification is
sometimes treated as an arbitrary choice to be made freely by the `observer'.
We object to this conventional view, holding that on the
contrary on any given physical situation there are only a few meaningful
ways to identify the relevant system, which are prescribed by the dynamics
and the limitations of observation. Let us clarify this important point,
as a way to approach our main concerns.

\subsection{Our approach: Decoherence of a nonlinear quantum field by
its own quantum fluctuations}

The criteria for choosing a particular subsystem for special treatment
(calling it relevant and the rest irrelevant \cite{ProjOp} is already
a preferential treatment), i.e., the definition of the open system,
is, to us, as important
a physical issue as finding the evolution of an open system itself. (For a
general discussion, see \cite{cgea,HuSpain}).
The possibility of successfully identifying a relevant system within a
complex physical problem hinges on the decoupling of some degrees of freedom
from the rest.
If the complete system is divisible into two sectors (subsystems) with
significant difference in their characteristic time, frequency, energy,
mass, length or interaction scales, then one can view one as the (open)
system and the other as the environment.
An example of mass discrepancy is the case
of quantum cosmology \cite{qc}, where the much heavier Planck mass makes it
possible to treat the gravitational sector differently via the
Born-Oppenheimer approximation. Decoherence of the `massive' gravitational
sector by the `lighter' matter field sector can lead to the emergence
of classical spacetimes in the semiclassical gravity regime \cite{decQC}.
Another example is the separation of `slow-fast' variables \cite{vanK}.
On the slow time scale, only the average action of the fast degrees of
freedom affects the relevant slow modes in an appreciable way. Factoring in
the asymptotic behavior of the fast variables, one can express their average
influence in terms of the slow variables themselves, thus obtaining an
effectively closed (and generally irreversible) dynamics for the latter
\cite{ProjOp}.

While there are many ways to split a complex system into a system proper and
an environment, only a few of these lead to physically interesting theories.
For example, not only the system proper should include everything of
interest to this or that particular observer, but also the dynamics of the
system proper should admit a closed, self-consistent description
(with some degree of stochasticity).
This requires that the system and bath to be weakly coupled, the system
being robust against the perturbation induced by its environment. Thus the
issue of the proper system - bath split in a definite situation is not to be
answered by the consideration of the observer's interests alone.  On the
contrary, the answer should be rooted in the physics of the system and the
observational context.
\footnote{There are more sophisticated ways to define an open system,
such as by the
partition of either physical or phase space into relevant and irrelevant
sectors (see, e.g., \cite{Hartle}).}

\subsubsection{Nonlinear fields and correlation dynamics}

As an open system is identifiably or dynamically separated from its
enviromment, decoherence occurs as it habitually interacts only with the
averaged environmental degrees of freedom {\it if and only if}
there are nontrivial correlations between the system and environment variables.
These correlations, in turn, may have a dynamical
origin, which requires nonlinear interactions between system and bath, or
else they may be present already in the initial conditions \cite{Balescu}.

Focusing on the correlational aspects, we have proposed earlier that a natural
way to partition a closed system is by way of the correlation functions,
defining the system as a subset of the BBGKY (classical) or Dyson (quantum)
hierarchies of correlation functions \cite{dch}.
The process of reducing the full dynamics to the autonomous
dynamics of the subsystem is usually described as
`truncation' and `factorization'.
In the Boltzmann molecular dynamics case this involves truncating
the BBGKY hierarchy at a correlation order
and assuming that this order of correlation function can be written as a
direct product of the lower order ones, known as the molecular
chaos assumption. The actual state of the environmental modes, however,
is never
quite equal to their truncated value; the small discrepancy is fed back into
the evolution of the system degrees of freedom as noise.
Effective autonomy of
the relevant system is needed for the stability and robustness which are
defining properties of classicality; but it is undermined constantly by the
effect of noise and fluctuations, which, as we have seen before \cite
{Tsukuba,GelHar2}, is instrumental to decoherence and the emergence of
classical behavior. Because
dissipation and noise are two aspects of the same underlying physics, they
are linked by consistency relations, known categorically as the
fluctuation - dissipation relations \cite{fdr},
which underly the theory of fluctuations in the stochastic, kinetic and
hydrodynamic regimes \cite{Landau}.
(Under equilibrium conditions, these relations
take the form of the famous Green - Kubo formulae \cite{fdr}.
The existence of such a relation in
non-equilibrium conditions is explored in \cite{nfsg,fdrsc}.)
Such is the necessary dynamical balance which prevails in
the quantum - classical interface.

In what follows, we shall  concentrate on nonlinear theories, and
seek to understand the generic conditions for a certain subset of the
degrees of freedom to decouple from the rest, while being decohered and
randomly driven by the remaining degrees of freedom.
Our goal is to show how these processes
actually occur in interacting field theories, and apply the results to
fluctuation generation in the early Universe which is only a manifestation
of this universal phenomenon.

\subsubsection{Mean field and quantum fluctuations}

In field theory applications, the different scales are usually associated to
the masses of the different particles, the mass being, in natural units, the
inverse correlation scale \cite{decoupling}. In the presence of spontaneous
symmetry breaking, another set of scales appear, associated with the
development of phase transition on one hand, and of quantum fluctuations
around the instantaneous value of the mean field on the other \cite{sdqft}.
Besides these, there is an intrinsic scale separation associated with
nonlinear quantum field theories, which arises because the physical,
observable excitations of the field are `dressed' by a cloud of virtual,
microscopic quantum fluctuations. Thus we can distinguish between the scale
associated with the physical or dressed excitations, and that associated
with the microscopic, elementary fluctuations \cite{CH88a}. The decoherence
of quantized dressed excitations is the main focus of this paper.

The description of the quantum to classical transition in terms of the
decoherence of the dressed field has several advantages over the
conventional procedure of splitting the field modes by hand into relevant
and irrelevant, the most important being that in this approach we do not
have to prejudge the importance of the different modes. Thus, for example,
in the actual application to fields in de Sitter space, we shall not require
any a priori consideration on the behavior of the different excitations on
horizon crossing. These considerations are sometimes hard to justify on a
rigorous basis, since the de Sitter horizon is an observer- dependent
construction, with no geometrical meaning. Moreover, our approach turns out
to be just the simplest of a hierarchy of increasingly accurate descriptions
of the field, where not only the dressed field but also other composite
operators are retained as relevant. We have presented the details of the full
approach elsewhere \cite{dch}.

\subsubsection{Organization of this paper}

In this paper we shall examine the process of decoherence of the dressed
field, and the corresponding development of a classical stochastic dynamics
for it, first on a simple example of a symmetry breaking theory, namely, a
scalar field theory in flat spacetime with a cubic self interaction, and
then in the physically relevant case of a free massless minimal field
propagating on a de Sitter background. This second model displays the basic
features of the fluctuation generation process in the early Universe though
in an elementary form.

The paper is organized as follows. In Sec. 2 we
discuss the decoherence of fluctuations in the dressed field of a self -
interacting, symmetry breaking field theory in flat space. For simplicity,
we shall only consider fluctuations around the false vacuum state, rather
than the phase transition in full generality. Our objective is to lay down the
basic elements of our approach, putting strong emphasis on the physical
processes linking dissipation, noise and decoherence to each other.
Sec. 3  applies the formalism above to a massless minimal field in
a de Sitter background. This can be taken as a simple model describing the
physics of fluctuations in the inflaton field in the early stage of inflation.
Despite its appearances, the theory
is nonlinear, because of the coupling of the scalar field to gravitons. Of
course, since we do not allow for correlations to be present in the initial
state, nonlinearity is a necessary condition for decoherence. The minimal
coupling of field and gravitation is the only nonlinear term which does not
detract from the predictive power of the model. In Sec. 4 we
discuss the main consequences of our findings.

A word about notations. We shall consider throughout a real scalar field in
flat space - time. The signature shall be $(-,+++)$. Fourier transforms are
defined as
\begin{equation}
\label{FT}A(x)\equiv \int ~{\frac{d^4k}{(2\pi )^4}}~e^{ikx}~A(k)
\end{equation}
where $kx=k_\mu x^\mu =-\omega t+\vec k.\vec x$. Always $\omega =k^0$. In
the case of translation-invariant kernels, we shall also define
\begin{equation}
\label{FT1}A(x,x^{\prime })\equiv \int ~{\frac{d^4k}{(2\pi )^4}}%
{}~e^{ik(x-x^{\prime })}~A(k)
\end{equation}
We shall assume that all interactions are adiabatically switched off in the
distant past, where the state of the field is the $IN$ vacuum. At finite
times, the state of the field will have evolved due to the influence of a
nontrivial background field. The notation $\langle \rangle \equiv \langle
IN||IN\rangle $ will always denote an expectation value with respect to this
state evolved from $IN$ vacuum. As a particular case, when the background
field vanishes the $IN$ vacuum persists; we shall identify expectation
values taken at zero background as $<>_0$.

In curved spacetime we shall use MTW conventions throughout \cite{MTW}. We
shall only consider fields on a fixed background de Sitter geometry, or
rather, that part of de Sitter space which can be described as a spatially
flat FLRW universe. In this case the role of $IN$ vacuum shall be filled by
the massless limit of the de Sitter invariant vacuum. Again, we shall use
$<> $ and $<>_0$ to denote expectation values at finite or vanishing
background field, respectively.

Later on, we shall have opportunity for computing variational derivatives of
various objects. The basic formula is

\begin{equation}
\frac{\delta \phi \left( x\right) }{\delta \phi \left( y\right) }=\delta
\left( x,y\right)
\end{equation}
where $\delta $ denotes the covariant Dirac distribution, defined from

\begin{equation}
\int d^4y\sqrt{-g\left( y\right) }\quad \delta \left( y,x\right) f\left(
y\right) =f\left( x\right)
\end{equation}

\section{Classical behavior of quantum fields}

In this Section we shall discuss the quantum to classical transition in
a nonlinear quantum field theory, taking as a working example the emergence of
classical stochastic behavior in a $g\phi ^3$ scalar field  in flat
space time. We shall adopt the consistent histories approach to quantum
physics \cite{conhis},
considering coarse-grained histories whose constitutive fine-grained
configurations are small departures from a given mean field. This
mean field may be interpreted as the physical quantum field, dressed by the
microscopic quantum fluctuations around it. We shall show that, in the limit
where the allowed variations of the field are small enough, the decoherence
functional is largely insensitive to the details of the ``window function''
defining the coarse graining procedure. Moreover, in this limit these
histories are consistent among themselves, in a sense to be made precise
below. The decoherence of these ``quantum mean field'' histories is closely
related to phenomena of noise and dissipation also present in the theory.

\subsection{Wave equation for fluctuations in the quantum mean field}

We consider a scalar field theory with action
\begin{equation}
\label{A1}S[\Phi ]=\int ~d^4x\{{-\frac 12}\partial _\mu \Phi \partial ^\mu
\Phi -V(\Phi )\}
\end{equation}
where the potential is
\footnote {Of course, more general forms are also possible, but this one is
convenient, for example, to study the onset of first order phase transitions.
For renormalization purposes, it is necessary to include a quartic term
as well;
we shall ignore this, assuming that the corresponding coupling constant
vanishes after all necessary subtractions have been carried out. See, e.g.
\cite{CH89}}
\begin{equation}
\label{B1}V(\Phi )=c\Phi +{\frac 12}m^2\Phi ^2-{\frac 16}g\Phi ^3
\end{equation}
The Heisenberg equations of motion are identical in form to the classical
field equation
\begin{equation}
\label{A3}-\Box \Phi (x)+{\frac{dV}{d\Phi (x)}}=0
\end{equation}
To identify the quantum mean field $\phi $, we write $\Phi =\phi +\varphi $,
where $\varphi $ represents small quantum fluctuations around $\phi $. Thus,
$\varphi $ obeys the linearized equation

\begin{equation}
\label{B1.2}-\Box \varphi +m^2\varphi -g\phi \varphi =0
\end{equation}
Subtracting this from the full equation, we find the equation for the
quantum mean field
\begin{equation}
\label{B1.1}-\Box \phi +c+m^2\phi -{\frac 12}g\phi ^2-{\frac 12}g\langle
\varphi ^2\rangle ={\frac 12}g(\varphi ^2-\langle \varphi ^2\rangle )
\end{equation}
where $<>$ denotes the vacuum expectation value of the $\varphi$ field, in
the background provide by the $\phi$ field. Later on, we shall use the
notation $<>_0$ to single out the expectation value computed at zero
background field.

Comparing  (\ref{B1.1}) with the original Heisenberg equation  (\ref
{A3}) we notice that the presence of the $\varphi $ field has modified the
inertia of the $\phi $ field. We interpret the solutions to  (\ref{B1.1})
as describing real excitations, which propagate surrounded by a cloud of
virtual $\varphi $ quanta. We shall consider the $\phi $ field as relevant,
and $\varphi $ as its environment. This procedure is meaningful insofar the
details of the $\varphi $ excitations are either irrelevant or inaccessible,
or both. In particular, we should be able to average out `fast' variables,
such as the phase of the $\varphi $ field, or equivalently, to assume that
these phases are actually random. In this regime, the right hand side of
(\ref{B1.1}) becomes small. If we drop it altogether, then  (\ref{B1.1})
admits a solution with $\phi \equiv 0$, the so - called `false vacuum',
provided
\begin{equation}
\label{B2}c={\frac 12}g\langle \varphi ^2\rangle _0={\frac 12}g\Delta
_F(x,x)
\end{equation}
where

\begin{equation}
\label{B5}\Delta _F(x,x^{\prime })=\int ~{\frac{d^4k}{(2\pi )^4}}~e^{ik(x-\
\acute x)}{\frac{-i}{(k^2+m^2-i\epsilon )}}
\end{equation}
is the Feynman propagator of quantum fluctuations around the false vacuum.
Henceforth, $\Delta $ will always denote the Green function of a microscopic
quantum fluctuation $\varphi $, that is, expectation values of binary
products of $\varphi $ field operators. In the following, we shall be
concerned as well with the propagation of perturbations in the dressed field
$\phi $ itself, which may also be described in terms of Green functions, for
which we shall use capital $G$ letters.

Assuming that $\phi $ remains small, we can linearize the left hand side of
 (\ref{B1.1}), to obtain the wave equation for the propagation of small
fluctuations in the quantum mean field. The right hand side is assumed to be
already small, and therefore is evaluated at the false vacuum value $\phi =0$
. Thus we obtain
\begin{equation}
\label{B1.3}-\Box \phi (x)+m^2\phi (x)-{\frac 12}g\int d^4x^{\prime }\frac{%
\delta \langle \varphi ^2\rangle (x)}{\delta \phi (x^{\prime })}\mid _{\phi
=0}\phi (x^{\prime })=gj(x)
\end{equation}
where

\begin{equation}
\label{C7.1}j(x)\equiv{\frac{1}{2}}\{\varphi^2(x)-\langle\varphi^2
\rangle_0(x)\}
\end{equation}
For latter use, let us call

\begin{equation}
\label{C17.2} {\frac{\delta \langle \varphi ^2\rangle (x)}{\delta \phi
(x^{\prime })}\mid _{\phi =0}} =-2gD(x,x^{\prime })
\end{equation}
and observe the elementary identity

\begin{equation}
\label{C2}D(x,x^{\prime })=[{\rm Im}(\Delta _F(x,x^{\prime }))^2]\theta
(t-t^{\prime })
\end{equation}
While we have the necessary data to compute this kernel explicitly, it is
actually more conducive for our purposes to observe that, because of Lorentz
invariance and the analytic properties associated with time ordering, the
square of the Feynman function admits a Lehmann representation
\begin{equation}
\label{C4}\Delta _F^2(x,x^{\prime })=-i\int ~{\frac{d^4k}{(2\pi )^4}}%
{}~e^{ik(x-x^{\prime })}\int_0^\infty {\frac{ds~h(s)}{(s+k^2-i\epsilon )}}
\end{equation}
where the function $h$ is positive and vanishes for $s\le s_0$, $s_0$ being
a positive threshold (an actual evaluation yields $h=\sqrt{1-(4m^2/s)}\theta
(s-4m^2)$). We find immediately
\begin{equation}
\label{C5}D(x,x^{\prime })={\frac 12}\int ~{\frac{d^4k}{(2\pi )^4}}%
{}~e^{ik(x-x^{\prime })}\int_0^\infty {\frac{ds~h(s)}{-(k+i\epsilon )^2-s}}
\end{equation}
(we use the notation $\left( k+i\epsilon \right) ^2=-(\omega +i\epsilon
)^2+\vec k^2$). We may now write down the equation of motion for the
fluctuations of the quantum mean field
\begin{equation}
\label{C7}\{-\Box _x+m^2\}\phi (x)+g^2\int ~d^4x^{\prime }~D(x,x^{\prime
})\phi (x^{\prime })=gj(x)
\end{equation}
where the right hand side is given by  (\ref{C7.1}). Obviously the
expectation value of the driving force vanishes, but its higher momenta do
not. In particular, we find
\begin{equation}
\label{C3}\langle \{j(x),j(x^{\prime })\}\rangle \equiv 2N(x,x^{\prime })=%
{\rm Re}(\Delta _F(x,x^{\prime }))^2
\end{equation}
More explicitly,
\begin{equation}
\label{C6}N(x,x^{\prime})={\frac{1}{2}}\int~{\frac{d^4k}{(2\pi)^4}}
{}~e^{ik(x-x^{\prime})}\pi h(-k^2).
\end{equation}

The kernel $N$, or rather, its Fourier transform ${\cal N}=h$, also plays a
distinguished role with respect to dissipation in this model. To begin with,
let us observe that, if
\begin{equation}
\label{C8}s_0+{\frac{g^2}{2}} \int~ds~{\frac{{\cal N}(s)}{(s-s_0)}}>m^2>
{\frac{g^2}{2}}\int~ds~{\frac{{\cal N}(s)}{s}}
\end{equation}
(which is largely satisfied within one loop accuracy) then this theory
admits stable one - particle asymptotic states of mass $M^2<s_0$, where

\begin{equation}
\label{C9}M^2+{\frac{g^2}{2}}\int~ds~{\frac{{\cal N}(s)}{(s-M^2)}}=m^2
\end{equation}

The properties of the quantum fluctuations of the mean field $\phi $ are
largely determined by the retarded propagator $G_r$, defined through its
Fourier components
\begin{equation}
\label{C11}G_r(k)=\left\{ [(k+i\epsilon )^2+m^2]-{\frac{g^2}{2}}
\int_0^{\infty} {\frac{ds~{\cal N}(s)}{s+(k+i\epsilon )^2}}\right\}^{-1}
\end{equation}
Or else, isolating the pole at $-k^2=M^2$,

\begin{equation}
\label{C12}G_r(k)={\frac{B}{[(k+i\epsilon )^2+M^2]}}+{\frac{g^2}{2}}
\int_0^{\infty}{\frac{ds~{\cal N}(s)}{((k+i\epsilon )^2+s)}} \vert
G_r(s)\vert^2
\end{equation}
where

\begin{equation}
\label{C14}B^{-1}=1+{\frac{g^2}2}\int ~ds~{\frac{{\cal N}(s)}{(s-M^2)^2}}
\end{equation}
and $G_r(s)$ means the propagator evaluated at any momentum $k$ with $k^2=-s$.
 (\ref{C12}) implies that a perturbation of the quantum mean field
propagates as an elementary free scalar field with mass $M^2$, superimposed
to a continuous spectrum of fields with masses ranging from $s_0$ to $\infty
$. The on-shell oscillations, with $k^2=-M^2$, are undamped. Above the $s_0$
threshold, however, oscillations are damped, as it can be seen from the $%
\phi $ field self energy developing a positive imaginary part. As we can see
from  (\ref{C11}), this imaginary part is again given by the function $%
h(-k^2)$. The same conclusion may be derived from the Feynman propagator for
the $\phi $ field, which, assuming vacuum initial conditions, is obtained
from the retarded propagator by analytical continuation $(k+i\epsilon )^2\to
(k^2-i\epsilon )$. As usual, this absorptive part in the field self-energy
is associated with the emission probability of real $\varphi $ quanta.
Indeed, we have shown in \cite{CH89} that the total amount of energy
dissipated from the quantum mean field is exactly the mean energy carried
away by the created particles.

The dual role of the $N$ kernel in both
fluctuation and dissipation, far from being an accident, follows from the
fluctuation - dissipation theorem. Indeed, always assuming vacuum initial
conditions, we may derive the Hadamard function for the quantum mean fields
(e. g., from the KMS condition at zero temperature \cite{KMS}) to be
\begin{equation}
\label{C18}G_1(k)=2\pi \{B\delta (k^2+M^2)+({\frac{g^2}2}%
)h(-k^2)|G_r(-k^2)|^2\}
\end{equation}

Since on-shell fluctuations are undamped, we may assume that the on - shell
contribution to the Hadamard function was already present in the initial
conditions for the mean field. However, such an interpretation would be
untenable above threshold, because there the $\phi $ field is damped, and
the memory of initial conditions is eventually lost. On the other hand, in
this part of the spectrum we have $\phi (k)=gG_r(k)j(k)$, so the
fluctuations in the driving force induce fluctuations in the mean field by
an amount
\begin{equation}
\label{C6.2}G_1(k)=2g^2|G_r(k)|^2{\cal N}(k)
\end{equation}
Comparing this with  (\ref{C18}) we conclude that the force self
correlation must indeed be given by  (\ref{C6}). The connection between
these fluctuations and particle creation is equally straightforward: While
dissipation describes the mean effects of particle creation, the source $j$
accounts for the deviation of the actual number of created particles from
this mean. The relationship between fluctuation and particle creation is
explored in full in Ref. \cite{nfsg}.

It is interesting to observe that the structure of the Hadamard kernel
(\ref{C18}) as the sum of on shell and off shell contributions, the latter
being related to dissipation, suggests that these fluctuations may be
regarded as independent. Should there be several decay channels for the
quantum mean field, then each would provide a further term to the Hadamard
function, so that the fluctuation - dissipation balance may hold.

\subsection{Decoherence of the Mean Field by its Quantum Fluctuations}

So far we have derived the wave equation for the quantum mean field $\phi $.
The equation of motion  (\ref{C7}) admits c-number solutions only under
the Hartree-Fock approximation $j\sim 0$. We now proceed to study under what
circumstances, if any, the dressed field is able to shed its quantum nature.
We adopt to this end the consistent histories approach to quantum mechanics
\cite{conhis}.

The basic tenet of this view of quantum mechanics is that quantum evolution
may be considered as a result of the coherent superposition of virtual
fine-grained histories, each carrying full information on the state of the
system at any given time. If we adopt the `natural' procedure of specifying
a fine grained history by defining the value of the field $\Phi (x)$ at
every spacetime point, these field values being c numbers, then the quantum
mechanical amplitude for a given history is $\Psi [\Phi ]\sim e^{S[\Phi ]}$
, where $S$ is the classical action evaluated at that particular history.
These histories are virtual because there exists interference between pairs
of histories. The strength of these effects is measured by the ``decoherence
functional''
\begin{equation}
\label{popa}{\cal D}[\Phi ,\Phi ^{\prime }]\sim \Psi [\Phi ]\Psi [\Phi
^{\prime }]^{*}\sim e^{i(S[\Phi ]-S[\Phi ^{\prime }])}
\end{equation}

On the other hand, our actual observations refer only to `coarse- grained'
histories, where several fine-grained histories are bundled together. A
coarse-grained history is defined, generally speaking, by a `filter
function' $\alpha $, which determines which fine-grained histories belong to
the superposition, and their relative phases. For example, we may have a
system with two degrees of freedom $x$ and $y$, and define a coarse-grained
history by specifying the values $x_0(t)$ of $x$ at all times. Then the
filter function is $\alpha [x,y]=\prod_{t\in R}\delta (x(t)-x_0(t))$. The
quantum mechanical amplitude for the coarse-grained history is defined as
\begin{equation}
\label{psiofalfa}\Psi [\alpha ]=\int ~D\Phi ~e^{iS}\alpha [\Phi ]
\end{equation}
We assume that the relevant information on the quantum state has been
encoded into the initial conditions for the paths in the integration domain.
The decoherence functional for two coarse-grained histories is \cite{conhis}
\begin{equation}
\label{dofalfa}{\cal D}[\alpha ,\alpha ^{\prime }]=\int ~D\Phi ^1D\Phi
^2e^{i(S(\Phi ^1)-S(\Phi ^2))}\alpha [\Phi ^1]\alpha ^{\prime }[\Phi ^2]^{*}
\end{equation}
The two histories $\Phi ^1$ and $\Phi ^2$ are not independent: they must
assume identical values on a $t=T={\rm constant}$ surface in the far future.
Decoherence means physically that the different coarse-grained histories
making up the full quantum evolution acquire individual reality, and may
therefore be assigned definite probabilities in the classical sense.
Therefore, as long as we remain within he accuracy afforded by the
coarse-graining procedure, we may disregard the quantum nature of our
system, and describe the dynamics as the self-consistent evolution of
c-number variables.

For our particular application, we wish to consider as a single coarse-
grained history all those fine- grained ones where the full field $\Phi $
remains close to a prescribed quantum mean field configuration $\phi $. Thus
the filter function $\alpha _\phi (\Phi )$ takes the form
\begin{equation}
\label{P1}\alpha _\phi (\Phi )=\int ~DJ~e^{iJ(\Phi -\phi )}\alpha _\phi (J)
\end{equation}
where $\alpha _\phi (J)$ is a smooth function (we explicitly exclude,
however, the case $\alpha _\phi \equiv {\rm constant}$, where there is no
coarse-graining at all). In  (\ref{P1}) we use the summation convention
over continuos indexes, i. e.
\begin{equation}
\label{R10}J\Phi \equiv \int ~d^4x~J(x)\Phi (x)
\end{equation}

The Decoherence Functional between two of these `mean field' histories is
then
\begin{equation}
\label{dofalfab} {\cal D}[\alpha_{\phi}, \alpha _{\phi ^{\prime}}]=\int~DJ
DJ^{\prime}~ e^{i\{ W[J,J^{\prime}]- (J\phi -J^{\prime}\phi^{\prime})\} }
\alpha_{\phi}[J]\alpha _{\phi ^{\prime}}[J^{\prime}]^*
\end{equation}
where

\begin{equation}
\label{R8}e^{iW[J,J^{\prime}]}=\int~D\Phi~D\Phi^{\prime}~e^{i(S[\Phi]-S^{*}[
\Phi^{\prime}]+ J\Phi-J^{\prime}\Phi^{\prime})}
\end{equation}
is precisely the closed-time-path (CTP) generating functional \cite{ctp}
. Since the filter functions are smooth, we may evaluate the integrals over
the $J$'s by saddle point methods, thus obtaining

\begin{equation}
\label{dfgamma}{\cal D}[\alpha _\phi ,\alpha _{\phi ^{\prime }}]={\cal C}%
(\phi ,\phi ^{\prime })e^{i\Gamma [\phi ,\phi ^{\prime }]}
\end{equation}
We recognize that $\Gamma $ is the closed-time-path effective action, and $%
{\cal C}$ is a slowly varying prefactor, namely
\begin{equation}
\label{prefac}{\cal C}(\phi ,\phi ^{\prime })\sim \alpha _\phi [-\Gamma
,_\phi ]\alpha _{\phi ^{\prime }}[\Gamma ,_{\phi ^{\prime }}]^{*}\{Det[{\
\frac{\delta ^2\Gamma }{\delta \phi ^a(x)\delta \phi ^b(x^{\prime })}}%
]^{1/2}\}
\end{equation}
where $a,b=1,2$, (e.g , $\phi ^1=\phi $, $\phi ^2=\phi ^{\prime }$). (\ref
{dfgamma}), which establishes the connection between the decoherence
functional for `mean field' histories and the closed-time-path effective
action, is a major result reported here. Of course, it is only particular
case of the more general ``correlation'' histories discussed in \cite{dch}.
For simplicity, we shall ignore the prefactor in what follows.

The evaluation of the closed time - path effective action is standard. To
one - loop accuracy it is given by \cite{CH87,CH89}
\begin{equation}
\label{A2}\Gamma [\phi ^a]=S[\phi ^a]+{\frac i2}{\rm lnDet}{\frac{\delta
^2S}{\delta \phi ^a\delta \phi ^b}}
\end{equation}
where $S[\phi ^a]$ is taken to mean $S[\phi ]-S[\phi ^{\prime }]^{*}$
(complex conjugation applies if an $i\epsilon $ term has been included to
enforce the boundary conditions), and the ``internal'' index $a$ is lowered
with the ``metric'' $g_{ab}={\rm diag}(1,-1)$. Functionally expanding $
\Gamma $ in powers of $\phi ^a$, and retaining only up to quadratic terms,
we get

\begin{equation}
\label{R14}\Gamma [\phi^a] ={\frac{1}{2}}\int~d^dxd^dx^{\prime}\{-[\phi ](x)
\tilde D(x,x^{\prime}) \{\phi\}(x^{\prime})+ig^2[\phi](x)N(x,x^{\prime})
[\phi ](x^{\prime})\}
\end{equation}
where $[\phi ]=(\phi^1-\phi^2)$, $\{\phi\}=(\phi^1+\phi^2)$,

\begin{equation}
\label{C1}\tilde D(x,x^{\prime })=\{-\Box _x+m^2\}\delta (x-x^{\prime
})+g^2D(x,x^{\prime })
\end{equation}
and the ``dissipation'' ($D$) and ``noise'' ($N$) kernels are defined in
Eqs. (\ref{C2}) and (\ref{C3}), respectively.

As discussed in the Introduction, fluctuations, namely, the presence of the
driving force  (\ref{C7.1}) on the right hand side of the wave equation
 (\ref{C7}) and decoherence, namely, the suppression of the decoherence
functional (\ref{dfgamma}) between different mean field histories both
depend on one and the same kernel $N$, related to the positive imaginary
part of the effective action, and are therefore revealed as aspects of the
same phenomenon. Note that both effects vanish if the cubic interaction is
switched off, revealing the essential role played by nonlinearity in this
problem. In turn, the presence of fluctuations is associated with the back
reaction of particle creation and thereby to dissipation: the two effects
are linked by the fluctuation- dissipation theorem. This manifests the
interrelation of decoherence, noise, and dissipation \cite{nfsg}, \cite{CM}
As have been shown earlier \cite{if}, the equations generated by the
effective action  (\ref{R14}) are equivalent to the linearized mean
field equations coupled to a stochastic Gaussian source $gj$, the noise
kernel $N$ being the auto-correlator of the source $j$. Comparing this with
the full equations (\ref{B1.3}), (\ref{C7.1}) we see that this equation lies
in between mean field theory, where the source is simply ignored, and the
full quantum theory. This approximation, moreover, successfully captures the
main property of the driving term, namely its mean square value  (\ref{C3}).
To fully account for non - Gaussian statistics, we must go to higher
loops and also include more complex correlation functions, employing the
more general methods described in \cite{dch}.

By repeating the arguments in the previous subsection, we see that the
mean-squared value of the decohered quantum mean field, as driven by the
stochastic source, is again given by (\ref{C6.2}). It is clear that this
amounts to only a fraction of the full quantum fluctuations, given by the
Hadamard function  (\ref{C18}). Thus, seeking the amount of classical
fluctuations subsequent to the quantum to classical transition by simply
equating the classical and quantum correlators, without a further analysis
of the decoherence process, is definitely unwarranted, unless it is meant as
a simple order of magnitude estimate. As shown elsewhere \cite{nfsg},
this fluctuation is related to the uncertainty in the number of created
particles from the dynamical quantum mean field.

Clearly, there is much more to be done to achieve a full understanding of
the quantum to classical transition in this model. For our present concerns,
however, we are satisfied with the observation that the quantum mean fields
may decohere through interaction with quantum fluctuations around them,
developing random classical fluctuations in the process. This phenomenon may
only occur in a nonlinear theory, and it is independent of any {\it a priori}
partition of fields or modes into relevant and irrelevant. Besides, there is
a rather powerful and comprehensive theory describing it, built with
well-proven techniques from non-equilibrium quantum field theory \cite
{if,ctp}. We shall now turn our attention to the problem of fluctuation
generation in the early Universe, to try and show it may be understood with
the same set of basic principles.

\section{Quantum Fluctuations and Density Perturbations in de Sitter Universe}

In this section we shall turn our attention to quantum fields in de Sitter
spacetime. Our goal is to describe, within the theory developed in the
previous section, how a quantum scalar field loses its quantum coherence,
and undergoes stochastic fluctuations in the process. During the
inflationary period, when the spacetime geometry can be approximated by the
de Sitter solution and the inflaton field described as a free, massles
scalar field, this may be seen as a model for the generation of primordial
fluctuations in the early Universe.

Here we shall seek a dynamical origin for decoherence (rather than imposing
a relevance criterium by hand). As we have seen in the previous section,
decoherence from an uncorrelated initial state can only occur in a nonlinear
theory. On the other hand, adding a self coupling to the inflaton field,
even leaving aside the stringent conditions imposed by the requirement of
`successful inflation', necessarily implies the inclusion of new parameters
into the model, making it correspondingly less compelling. Therefore we are
led to consider the only available parameter-free source of nonlinearity,
namely, the gravitational couplings of the inflaton. To appeal to quantum
effects of the gravitational field immediately evokes a number of
difficulties arising from the non- renormalizability of general relativity.
In this work we shall sidestep this issue, by considering only one loop
effects. Moreover, as in the previous section, we shall not carry through
the renormalization procedure explicitly, but rather assume that the theory
has already been rendered finite by adding suitable counterterms to the
classical action.
In fact, we shall base our analysis on the Einstein - Hilbert
form of the action, without including higher order terms which could arise
in the renormalization process. This procedure is fully justified at the
scales of interest \cite{donoghue}. Another feature of quantum gravity which
we shall sidestep is the gauge character of the gravitational field. To
highlight the physical ideas in our approach, we shall take the simple-
minded view of fixing the gauge at the classical level, considering only
quantum fluctuations of the `physical' degrees of freedom \cite{fordparker}.
We shall present the results of a more complete calculation elsewhere.

Thus we shall consider a theory involving two quantum fields, the
gravitational field $g_{\mu \nu }^q$ and the inflaton field $\Phi $. The
classical action functionals are given respectively by
\begin{equation}
\label{sg1}S_g=m_p^2\int d^4x\sqrt{-g^q}\left\{ R^q-2\Lambda \right\}
\end{equation}
and
\begin{equation}
\label{sf1}S_f=-\frac 12\int d^4x\sqrt{-g^q}\partial _\mu \Phi \partial ^\mu
\Phi \
\end{equation}

As in the previous Section, we shall consider coarse-grained histories
defined by the values of the quantum mean fields $g_{\mu \nu }$ and $\phi $.
The decoherence functional is related to the CTP effective action as in  (%
\ref{dfgamma} ). Following the usual prescription for the computation of the
one-loop effective action, we write both fields in terms of the quantum mean
(or dressed) fields and their fluctuations:
\begin{equation}
\label{gq1}g_{\mu \nu }^q=g_{\mu \nu }+h_{\mu \nu },
\end{equation}
and
\begin{equation}
\label{phiq1}\Phi =\phi +\varphi .
\end{equation}
To make sure that the fluctuations are physical, we work with the transverse
traceless gauge in  synchronous coordinates, namely,

\begin{equation}
\label{gaugecond}h_\mu ^\mu =h_{\mu ;\nu }^\nu =0;~~h_\mu ^0=0
\end{equation}
where indices are raised and lowered with the background metric, and the
derivative is taken with the background Levi - Civita connection \cite
{fordparker}. Observe that the classical equations of motion admit a
solution with
\begin{equation}
\label{desitter}g_{\mu \nu }=(H\tau )^{-2}\eta _{\mu \nu }
\end{equation}
and with vanishing field. Here $\tau \leq 0$, where $\tau \equiv \int dt/a$
is the conformal time, $\Lambda =3H^2$, and $\eta _{\mu \nu }$ is the flat
space time metric. ( This solution, of course, represents only one half of
de Sitter space time \cite{schrodinger}.) Since we are not concerned at this
moment with the stochastic fluctuations of the gravitational field itself
(see \cite{nfsg} in this connection), we shall compute the noise and
dissipation kernels for this value of the gravitational background, leaving
only $\phi $ arbitrary.

Continuing with the computation of the dissipative and stochastic elements
in the dynamics of the inflaton, we should expand the classical action in
powers of the perturbations, and retain terms only up to quadratic orders.
This is of course equivalent to computing the full nonlinear equation, and
linearizing afterwards, as we did in Section 2, but in a more complex
theory, this approach is more efficient. Beginning with the scalar action
for simplicity, we obtain three kinds of terms which are independent of,
linear and quadratic in $\varphi $, respectively. The term which does not
contain $\varphi $ is necessarily quadratic in both $h_\nu ^\mu $ and $\phi $.
To one loop accuracy it only appears in `tadpole' graphs, with no
relationship to the non local part of the noise and dissipation kernels, and
we will not consider it further. The part quadratic in $\varphi $ defines
the propagator for these microscopic fluctuations. It takes the form
\begin{equation}
\label{sf2}S_f^{(2)}=\frac 12\int d\tau d^3x(H\tau )^{-2}\left( \varphi
^{\prime }-\vec \nabla \varphi ^2\right) ,
\end{equation}
where the prime stands for a $\tau $ derivative. The $\varphi h$ cross term
is the source of dissipation, decoherence, and noise in this model. For
physical gravitational perturbations (that is, those obeying $%
h_0^0=h_i^0=h_i^i=h_{i,j}^j=0$), it is given by
\begin{equation}
\label{sint}S_f^{\left( 1,1\right) }=-\int d\tau \;d^3x\;(H\tau )^{-2}\phi
_{,ij}\;h_i^j\varphi ,
\end{equation}
where Latin indexes run from 1 to 3.

Expanding the Einstein - Hilbert action we can read out the free graviton
propagator. The quadratic terms in the action are
\begin{equation}
\label{sg2}S_g^{(2)}=\frac{m_p^2}4\int d\tau \;d^3x(H\tau )^{-2}(h_j^{\prime
i}h_i^{\prime j}-h_{j,k}^ih_{i,k}^j).
\end{equation}
However, the graviton components $h_j^i$ are not independent, since they are
linked through the gauge conditions. It is convenient to write the graviton
field explicitly in terms of the independent physical degrees of freedom, as
in \cite{fordparker}
\begin{equation}
h_j^i\left( \tau ,\vec x\right) =\left( \frac 1{m_p}\right) \int d^3y\left\{
G_j^{+i}\left( \vec x-\vec y\right) h^{+}\left( \tau ,\vec y\right)
+G_j^{\times i}\left( \vec x-\vec y\right) h^{\times }\left( \tau ,\vec
y\right) \right\}
\end{equation}
where

\begin{equation}
G_j^{+i}\left( \vec x-\vec y\right) =\int \frac{d^3k}{\left( 2\pi \right) ^3}%
e^{i\vec k(\vec x-\vec y)}A_{\vec kj}^{+i}
\end{equation}
and a similar formula exists for the cross ($\times $) polarization. The $A$
matrices obey
\begin{equation}
A_{\vec ki}^{+i}=k_iA_{\vec kj}^{+i}=A_{\vec ki}^{\times i}=k_iA_{\vec
kj}^{\times i}=0
\end{equation}

\begin{equation}
A_{\vec kj}^{+i}A_{-\vec kl}^{+j}=A_{\vec kj}^{\times i}A_{-\vec kl}^{\times
j}=\delta _l^i-\frac{k^ik_l}{k^2}
\end{equation}
\begin{equation}
A_{\vec kj}^{+i}A_{-\vec ki}^{\times j}=0
\end{equation}

The graviton action thus splits into two parts,
\begin{equation}
S_g^{(2)}=S^{+}+S^{\times }
\end{equation}
each being the action for a massless, real scalar field,  (\ref{sf2}). We
also find

\begin{equation}
S_f^{\left( 1,1\right) }=-(\frac 1{m_p})\int \frac{d\tau \;d^3xd^3y}{\left(
H\tau \right) ^2}\;\left( \phi _{,ij}\varphi \right) \left( \tau ,\vec
x\right) \left\{ G_j^{+i}\left( \vec x-\vec y\right) h^{+}\left( \tau ,\vec
y\right) +G_j^{\times i}\left( \vec x-\vec y\right) h^{\times }\left( \tau
,\vec y\right) \right\}
\end{equation}
While it is possible to derive the effective action for this model, in order
to find the noise and dissipation kernels it is simplest to proceed from the
equations of motion, as given in Section 2.1. Let us begin with the
Heisenberg equation of motion for the inflaton field
\begin{equation}
\Box \Phi -\frac{\left( H\tau \right) ^2}{m_p}\partial _{ij}^2\int
d^3y\;\varphi \left( \tau ,\vec x\right) \left\{ G_j^{+i}\left( \vec x-\vec
y\right) h^{+}\left( \tau ,\vec y\right) +G_j^{\times i}\left( \vec x-\vec
y\right) h^{\times }\left( \tau ,\vec y\right) \right\} =0
\end{equation}
It is clear that in the absense of a nontrivial background field, the
expectation value
\begin{equation}
\left\langle \;\varphi \left( \tau ,\vec x\right) h^{+,\times }\left( \tau
,\vec y\right) \right\rangle \mid _{\phi =0}\equiv 0
\end{equation}
Thus the linearized equation for the mean field reads
\begin{equation}
\label{meanfielddyn}\Box \phi \left( x\right) -\int \frac{d^4x^{\prime }}{%
\left( H\tau ^{\prime }\right) ^4}D\left( x,x^{\prime }\right) \phi \left(
x^{\prime }\right) =0
\end{equation}
where
\begin{equation}
\label{disker}\;D\left( x,x^{\prime }\right) =\frac{\left( H\tau \right) ^2}{%
m_p}\partial _{ij}^2\int d^3y\;\left\{ G_j^{+i}\left( \vec x-\vec y\right)
\frac \delta {\delta \phi (x^{\prime })}\left\langle \;\varphi \left( \tau
,\vec x\right) h^{+}\left( \tau ,\vec y\right) \right\rangle \mid _{\phi
=0}+\left( +\leftrightarrow \times \right) \right\}
\end{equation}

Comparing the mean field equation with the Heisenberg equation, we obtain
the equation for the dressed fluctuations
\begin{equation}
\label{fullmeanfield}\Box \phi \left( x\right) -\int \frac{d^4x^{\prime }}{%
\left( H\tau ^{\prime }\right) ^4}D\left( x,x^{\prime }\right) \phi \left(
x^{\prime }\right) =\frac{\left( H\tau \right) ^2}{m_p}j(x)
\end{equation}
where in principle $j$ represents the composite operator
\begin{equation}
j\left( x\right) =\int d^3y\;\partial _{ij}^2\varphi \left( \tau ,\vec
x\right) \left\{ G_j^{+i}\left( \vec x-\vec y\right) h^{+}\left( \tau ,\vec
y\right) +G_j^{\times i}\left( \vec x-\vec y\right) h^{\times }\left( \tau
,\vec y\right) \right\}
\end{equation}
(Here we have used the transversal character of the $G$ tensors). As we
discussed in detail in Section 2, upon decoherence we can think of $j$ as a
classical stochastic source, whose self correlation is given by the noise
kernel
\begin{equation}
\label{noiseker}N\left( x,x^{\prime }\right) =\left\langle j\left( x\right)
j\left( x^{\prime }\right) \right\rangle _c=\frac 12\left\langle \left\{
j\left( x\right) ,j\left( x^{\prime }\right) \right\} \right\rangle _0
\end{equation}

To compute the quantum expectation values in Eqs. (\ref{disker}) and (\ref
{noiseker}), we expand the quantum field operators in terms of the
destruction and creation operators, as in
\begin{equation}
\label{modedecomp}\varphi \left( \tau ,\vec x\right) =iH\int \frac{d^3k}{
\left( 2\pi \right) ^{3/2}k}\frac{e^{i\vec k\vec x}}{\sqrt{2k}}\left\{ a_{\vec
k}f_k\left( \tau \right) +a_{-\vec k}^{\dagger }f_k^{*}\left( \tau \right)
\right\}
\end{equation}
and perform similar expansions for the graviton amplitudes. All three scalar
fields $\varphi $, $h^{+}$, and $h^{\times }$ are expanded in terms of the
same modes

\begin{equation}
f_k\left( \tau \right) =e^{-ik\tau }\left[ 1+ik\tau \right]
\end{equation}
which are related to the Fulling - Davies vacuum, this being the natural
choice of quantum state in this problem \cite{cosmicbaldness}. The Wronskian
of these modes is

\begin{equation}
\label{wronskian}f_k^{*}\left( \tau \right) f_k^{\prime }\left( \tau \right)
-f_k\left( \tau \right) f_k^{*^{\prime }}\left( \tau \right) =-2ik^3\tau ^2
\end{equation}

\subsection{Dissipation and dynamics of free mean fields}

Before continuing with the discussion of the fluctuations in the
quantum mean field,
we want to first analyze the solutions to the source-free
mean field equation,  (\ref{meanfielddyn}). Concretely, our goal is to
stablish the dissipative character of this equation, to be able later to
analyze the fluctuations in terms of the fluctuation - dissipation relation.

As we show in the Appendix, the dissipation kernel is conveniently written as

\begin{equation}
\label{disker2}D\left( x,x^{\prime }\right) =\frac{H^4}{m_p^2}\theta \left(
\tau -\tau ^{\prime }\right) \left( H\tau \right) ^2\left( H\tau ^{\prime
}\right) ^2\int \frac{d^3k}{\left( 2\pi \right) ^3}e^{i\vec k\left( \vec
x-\vec x^{\prime }\right) }D_k\left( \tau ,\tau ^{\prime }\right)
\end{equation}
where

\begin{equation}
\label{disker3}D_k\left( \tau ,\tau ^{\prime }\right) =\frac 1{\left( 2\pi
\right) ^3}\int \frac{d^3p}{2p^3}\frac{d^3q}{2q^3}\delta \left( \vec p+\vec
q-\vec k\right) \Theta \left( \vec p,\vec q\right) F_{pq}\left( \tau ,\tau
^{\prime }\right)
\end{equation}
and
\begin{equation}
\label{theta}\Theta (\vec p,\vec q)=p_ip_jp_lp_m\left[ A_{\vec
qj}^{+i}A_{-\vec ql}^{+j}+A_{\vec qj}^{\times i}A_{-\vec ql}^{\times j}\right],
\end{equation}

\begin{equation}
\label{fpq}F_{pq}\left( \tau ,\tau ^{\prime }\right) =2{\rm Im}\left[
f_p^{*}\left( \tau \right) f_q^{*}\left( \tau \right) f_p(\tau ^{\prime
})f_q(\tau ^{\prime })\right].
\end{equation}

Since the background is spatially homogeneous, we can also expand the quantum
mean field in terms of its Fourier modes:

\begin{equation}
\label{fourierdec}\phi \left( \tau ,\vec x\right) =H\int \frac{d^3k}{\left(
2\pi \right) ^{3/2}k}\frac{e^{i\vec k\vec x}}{\sqrt{2k}}\phi _k\left( \tau
\right)
\end{equation}
The time dependent amplitude $\phi _k\left( \tau \right) $ can always be
written as

\begin{equation}
\phi _k\left( \tau \right) =\alpha _{\vec k}f_k\left( \tau \right) +\beta
_{\vec k}f_k^{*}\left( \tau \right).
\end{equation}
Imposing the auxiliary condition

\begin{equation}
\phi _k^{\prime }\left( \tau \right) =\alpha _{\vec k}f_k^{\prime }\left(
\tau \right) +\beta _{\vec k}f_k^{\prime *}\left( \tau \right),
\end{equation}
and keeping the Wronskian  (\ref{wronskian}) in mind, we find

\begin{equation}
\alpha _k\left( \tau \right) = \frac i{2k^3} \left[
f_k^{*}\left( \tau \right) \left( \frac{\phi _k^{\prime }\left( \tau \right)
}{\tau ^2}\right) -\phi _k\left( \tau \right) \left( \frac{f_k^{*^{\prime
}}\left( \tau \right) }{\tau ^2}\right) \right]
\end{equation}

\begin{equation}
\beta _k\left( \tau \right) = - \frac{i}{2k^3} \left[ f_k\left(
\tau \right) \left( \frac{\phi _k^{\prime }\left( \tau \right) }{\tau ^2}%
\right) -\phi _k\left( \tau \right) \left( \frac{f_k^{\prime }\left( \tau
\right) }{\tau ^2}\right) \right]
\end{equation}

In the absense of dissipation, the coefficients $\alpha $ and $\beta $ would
be constant. In the presence of the dissipation kernel, they become
functions of time, with evolution equations

\begin{equation}
\label{alfapr}\tau ^2\alpha '_k\left( \tau \right) = \frac i{2k^3}
 \frac{H^4}{m_p^2}f_k^{*}\left( \tau \right) \int^\tau \frac{%
d\tau ^{\prime }}{\left( H\tau ^{\prime }\right) ^2}D_k\left( \tau ,\tau
^{\prime }\right) \phi _k\left( \tau ^{\prime }\right)
\end{equation}

\begin{equation}
\label{betapr}\tau ^2\beta '_k\left( \tau \right) =-\frac{i}{2k^3}
 \frac{H^4}{m_p^2}f_k\left( \tau \right) \int^\tau \frac{d\tau
^{\prime }}{\left( H\tau ^{\prime }\right) ^2}D_k\left( \tau ,\tau ^{\prime
}\right) \phi _k\left( \tau ^{\prime }\right)
\end{equation}
We are interested in the solution where $\alpha \rightarrow 1$ and $\beta
\rightarrow 0$ in the far past. In a first approximation, we may substitute
$\phi _k$ by its free value $f_k$ in Eqs.(\ref{alfapr}) and (\ref{betapr}).
In particular, we obtain, for the total integrated change in the amplitude of
the mode $\sigma _k\equiv \left| \alpha _k\right| ^2$ ,

\begin{equation}
\Delta \sigma _k=-\frac{2H^6}{m_p^2k^3}\int \frac{d\tau }{\left( H\tau
\right) ^2}\frac{d\tau ^{\prime }}{\left( H\tau ^{\prime }\right) ^2}\theta
\left( \tau -\tau ^{\prime }\right) D_k\left( \tau ,\tau ^{\prime }\right)
{\rm Im}\left[ f_k\left( \tau ^{\prime }\right) f_k^{*}\left( \tau \right)
\right]
\end{equation}
Substituting the value of $D_k$, this yields

\begin{equation}
\label{deltasigma}\Delta \sigma _k=-\frac{H^2}{m_p^2\left( 2\pi \right) ^3k^3%
}\int \frac{d^3p}{2p^3}\frac{d^3q}{2q^3}\delta \left( \vec p+\vec q-\vec
k\right) \Theta \left( \vec p,\vec q\right) \left\{ \left| F_{kpq}\right|
^2-\left| G_{kpq}\right| ^2\right\}
\end{equation}
where

\begin{equation}
\label{fkpq}F_{kpq}=\int \frac{d\tau }{\tau ^2}f_k\left( \tau \right)
f_p^{*}\left( \tau \right) f_q^{*}\left( \tau \right)
\end{equation}

\begin{equation}
G_{kpq}=\int \frac{d\tau }{\tau ^2}f_k\left( \tau \right) f_p\left( \tau
\right) f_q\left( \tau \right).
\end{equation}

Clearly, the more rapidly oscillatory function $G$ will be much smaller
than $F$, and we
shall neglect it in what follows. It should be observed, besides, that when
$k=p+q$, the condition $\vec k=\vec p +\vec q$, enforced by the delta
function in  (\ref{deltasigma}), implies that $\vec p$ and $\vec q$
are colinear, in which case $\Theta  \left( \vec p,\vec q\right) =0$.
Because of this, the integrand in $F_{kpq}$ is always oscillatory, and the
integral is independent of the lower limit of integration.

 To summarize, if we ignore the effects of
fluctuations, we must conclude that the mean field is dissipated by its
interaction with the environment, losing an amount $\Delta \sigma _k$ of its
original amplitude (and an equal amount of its original Klein Gordon charge)
over the de Sitter period of cosmic evolution. Since on the other hand zero
point fluctuations cannot disappear, we should expect that an equal amount
will be provided by the environment, now under the guise of random driving
force, so that the fluctuation - dissipation balance may be kept.
We turn now to investigate this issue.

\subsection{Noise and the fluctuation - dissipation balance}

Let us return to the full dynamics of the quantum mean field,
as described by  (\ref{fullmeanfield}).
We see that, besides the dissipative terms just
analyzed, the field is coupled to a random source $j$, whose mean square
value is given by the noise kernel (\ref{noiseker}). Substituting the mode
decomposition (\ref{modedecomp}), it is straightforward to find the
explicit expression

\begin{equation}
\label{noiseker2}N\left( x,x^{\prime }\right) =H^4\int \frac{d^3k}{\left(
2\pi \right) ^32k^3}\frac{d^3p}{\left( 2\pi \right) ^32p^3}e^{i\left( \vec
k+\vec p\right) \left( \vec x-\vec x^{\prime }\right) }\Theta (\vec k,\vec p)%
{\rm Re}\left[ f_k\left( \tau \right) f_p\left( \tau \right) f_k^{*}(\tau
^{\prime })f_p^{*}(\tau ^{\prime })\right]
\end{equation}

More interesting than the noise kernel are the fluctuations induced on the
quantum mean field $\phi $ itself. From the point of view of the theory of
primordial fluctuations in the Universe, the most relevant quantity is the
mean square value of the fluctuations at a given time, which are given by
\begin{equation}
\label{fluct}<\phi (\tau ,\vec x)\phi (\tau ,0)>_c=\frac 1{m_p^2}\int \frac{%
d^4r_1}{\left( H\tau _1\right) ^2}\frac{d^4r_2}{\left( H\tau _2\right) ^2}%
G_r\left( \left( \tau ,\vec x\right) ,r_1\right) G_r\left( \left( \tau
,0\right) ,r_2\right) N(r_1,r_2)
\end{equation}
In a first approximation, we may use free retarded propagators for the $G$
Green functions
\begin{equation}
G_r\left( x,x_1\right) =-iH^2\theta \left( \tau -\tau _1\right) \int \frac{%
d^3k}{\left( 2\pi \right) ^32k^3}e^{i\vec k(\vec x-\vec x_1)}\left\{
f_k\left( \tau \right) f_k^{*}\left( \tau _1\right) -f_k\left( \tau
_1\right) f_k^{*}\left( \tau \right) \right\}
\end{equation}
The result is

\begin{equation}
<\phi (\tau ,\vec x)\phi (\tau ,0)>_c=\int d^3k\;e^{i\vec k\vec x}\;\left|
\delta \phi _k\right| ^2\left( \tau \right)
\end{equation}
where

\begin{equation}
\left| \delta \phi _k\right| ^2= \frac {H^2}{\left( 2\pi \right)
^32k^3} F\left( \tau ,k\right)
\end{equation}
and
\begin{equation}
F\left( \tau ,k\right) = \frac {H^2}{ (2\pi)^3m_p^2k^3} \int \frac{d^3q}{2q^3}
\frac{d^3p}{2p^3}\delta \left( \vec k-\vec p-\vec q\right) \Theta (\vec
q,\vec p)\mid \left\{ f_k\left( \tau \right) F_{kpq}^{*}\left( \tau \right)
-f_k^{*}\left( \tau \right) G_{kpq}^{*}\left( \tau \right) \right\} \mid ^2
\end{equation}
($\Theta $ was defined in  (\ref{theta}), $F$ and $G$ in  (\ref
{deltasigma})). If, as in the previous subsection, we neglect $G$ compared to
$F$, this reduces to the simple result

\begin{equation}
\label{fdt}F\left( \tau ,k\right) =-\Delta \sigma _k\left| f_k\left( \tau
\right) \right| ^2
\end{equation}

This result, of course, is exactly what we should expect from the fluctuation -
dissipation arguments. The environment injects into the system exactly the
amount of fluctuations necessary to mantain consistency with equilibrium
against
the tendency of the mean field to dissipate away. We could say that
the environment returns as noise what it had previously absorbed as
dissipation; the important point is that in the process  these fluctuations
have been degraded from coherent quantum fluctuations to incoherent
stochastic ones. This reprocessing of part of the quantum mean field by the
environment is the physical content of the decoherence process.

An important consequence of  (\ref{fdt}) is that only a fraction, and
indeed a very small part, of the total zero point fluctuations may ever
become classical, and thus contribute to structure formation, unless the
Hubble parameter $H$ be of the order of the Planck mass. This is the crucial
point relevant to the  cosmological problem, and therefore
deserves to be elaborated in some detail.

\section{Generation of fluctuations in inflationary cosmology}

So far we have presented a comprehensive framework to study the
transmutation of quantum into classical fluctuations in nonlinear field
theories, and applied it to a scalar field propagating on a de Sitter
background while interacting with gravitons.
Let us now apply these results to the problem of the generation of
primordial fluctuations during the inflationary era. We shall contrast
the well-accepted results in the literature with that obtained by our method.

For simplicity, we follow Guth and Pi's treatment \cite{guthpi} for
the density contrast derived from quantum fluctuations in the inflaton field.
It is estimated to be
\begin{equation}
{\frac{{\delta \rho }}\rho }\approx {\frac{{H\delta \phi _k}}{{<\dot \Phi >}}}
\end{equation}
where the right hand side is evaluated at the time a given mode
`leaves' the de Sitter horizon,  i.e., when $k\tau \sim 1$. As a
viable example, we shall adopt the `chaotic' inflation \cite{ChaoInf}
model, where the inflaton field self- interacts
with a $\lambda \Phi ^4$ effective potential. During the inflationary
period, the vacuum energy dominates the stress energy tensor of the field,
and the inflaton slowly rolls down the potential well. Because of the first
assumption, the Hubble parameter becomes

\begin{equation}
H^2\sim \frac{\lambda \Phi ^4}{m_p^2}
\end{equation}
(in this discussion, we shall systematically ignore factors of order unity).
Because of the second assumption, the equation of motion for the field is

\begin{equation}
\dot \Phi +\sqrt{\lambda }m_p\Phi =0
\end{equation}
with solution

\begin{equation}
\Phi \left( t\right) =\Phi _0e^{-\sqrt{\lambda }m_pt},
\end{equation}
and
\begin{equation}
H\left( t\right) =\frac{\sqrt{\lambda }\Phi _0^2}{m_p}e^{-2\sqrt{\lambda }%
m_pt}
\end{equation}
where we have placed the origin of cosmic proper time at the beginning of
inflation. Both assumptions break down when $\Phi \sim m_p$, so we estimate
the length of the inflationary period $\Delta t$ as

\begin{equation}
e^{\sqrt{\lambda }m_p\Delta t}\sim \frac{\Phi _0}{m_p}
\end{equation}
and the number of e-foldings

\begin{equation}
n=\int^{\Delta t}Hdt\sim \left( \frac{\Phi _0}{m_p}\right) ^2.
\end{equation}
A satisfactory resolution of the horizon problem demands $n\geq 60$.
This implies that the variation of $H$ over an e-folding is small.

With these inputs, we may compute the spectrum of primordial fluctuations. In
the conventional treatment, where the full quantum fluctuations of the
inflaton are seen as contributing to structure formation,  $\delta \phi _k$
is read directly out of the mode expansion as \cite{vilenkin}

\begin{equation}
\left| \delta \phi _k\right|^2 _{{\rm naive}}= \frac {H^2}{\left( 2\pi
\right) ^32k^3}
\end{equation}
Therefore

\begin{equation}
\frac{\delta \rho }\rho \sim \frac{H^2}{\dot \Phi }\sim \sqrt{\lambda }%
\left( \frac \Phi {m_p}\right) ^3.
\end{equation}
The physically most relevant modes are those which leave the horizon late in
the inflationary period, when $\Phi \sim m_p$. For these modes, the
observational constraint $\delta \rho /\rho \sim 10^{-6}$ at decoupling
leads to a severe bound on the inflaton interactions $\lambda \sim 10^{-12}
$ . This is one of the outstanding puzzles in inflationary cosmology.

If we compare our results for the semiclassical fluctuations with the
usual estimates in the literature, we find they differ by the presence of
the $F$ factor. Closer examination reveals that the integral defining $%
F\left( \tau ,k\right) $   depends on its arguments only through the
combination $k\tau $, and as the mode `leaves the horizon' it becomes a
dimensionless constant. ( Of course, if we take the defining expression at
face value,  this constant would be infinite, but, since the divergence is
only logarithmic, after suitable ultraviolet and infrared cutoffs are
introduced, the physical result shall be of order one.). Therefore we simply
obtain

\begin{equation}
F\sim \left( \frac H{m_p}\right) ^2\sim \lambda \left( \frac \Phi
{m_p}\right) ^4
\end{equation}
And, for `short' wavelength modes,

\begin{equation}
{\frac{{\delta \rho }}\rho \sim \lambda }
\end{equation}

This correction modifies the above bound on $\lambda $ by six orders of
magnitude, i. e., we have $\delta \rho /\rho \sim 10^{-6}$, with
$\lambda \sim 10^{-6}$. This represents a dramatic reduction in the fine
tuning required by the model; in fact, this value of $\lambda $ is
consistent with the inflaton taking part in nonabelian gauge interactions
with a coupling constant of $10^{-2}$, while the older estimate would
require to shield the inflaton unnaturally from radiative corrections. On
the other hand, the value of $\lambda $ is not so high as to make the
coupling of the inflaton with its own fluctuations prevail over the
gravitational couplings considered here.

As we have illustrated in this paper, with proper consideration of
decoherence and noise for quantum fields, the possibility of
developing successful inflationary scenarios within moderately
nonlinear field theories has far-reaching consequences. Not only can we
place the inflaton field in the proper ranges of conventional high energy
physics in the treatment of fluctuations, but also better implement
the inflaton dynamics \cite{HuRav}, entropy generation \cite{KMHR}
and reheating problems \cite{RSH}.
We are continuing this line of investigation on these outstanding issues.
\section{Appendix}

\subsection*{A. Dissipation Kernel}

Let us begin with the calculation of the dissipation kernel  (\ref
{disker}). In order to compute the variational derivatives, we must
consider the equations for the $\varphi $ and $h$ fields, namely
\begin{equation}
\Box \varphi \left( x\right) =\frac{\left( H\tau \right) ^2}{m_p}\phi
_{,ij}\left( x\right) \int d^3z\left\{ G_j^{+i}\left( \vec x-\vec z\right)
h^{+}\left( \tau ,\vec z\right) +\left( +\leftrightarrow \times \right)
\right\}
\end{equation}
\begin{equation}
\Box h^{+}\left( \tau ,\vec y\right) =\frac{\left( H\tau \right) ^2}{m_p}%
\int d^3z\left\{ G_j^{+i}\left( \vec y-\vec z\right) \left[ \phi
_{,ij}\varphi \right] \left( \tau ,\vec z\right) \right\}
\end{equation}
and a similar equation for the cross ($\times $) polarization . Taking
variational derivatives of both sides of these equations, we find

\begin{equation}
\Box \frac{\delta \varphi \left( x\right) }{\delta \phi \left( x^{\prime
}\right) }\mid _{\phi =0}=\frac{\left( H\tau \right) ^2}{m_p}\partial
_{ij}^2\delta \left( x,x^{\prime }\right) \int d^3z\left\{ G_j^{+i}\left(
\vec x^{\prime }-\vec z\right) h^{+}\left( \tau ,\vec z\right) +\left(
+\leftrightarrow \times \right) \right\}
\end{equation}

\begin{equation}
\Box \frac{\delta h^{+}\left( \tau ,\vec y\right) }{\delta \phi \left(
x^{\prime }\right) }\mid _{\phi =0}=\frac{\left( H\tau ^{\prime }\right) ^2}{%
m_p}\delta \left( \tau ,\tau ^{\prime }\right) G_j^{+i}\left( \vec y-\vec
x^{\prime }\right) \partial _{ij}^2\varphi \left( x^{\prime }\right)
\end{equation}

These equations may be solved as

\begin{equation}
\frac{\delta \varphi \left( x\right) }{\delta \phi \left( x^{\prime }\right)
}\mid _{\phi =0}=\frac{\left( H\tau ' \right) ^2}{m_p}\partial _{ij}^2\Delta
_{ret}\left( x,x^{\prime }\right) \int d^3z\left\{ G_j^{+i}\left( \vec
x^{\prime }-\vec z\right) h^{+}\left( \tau ,\vec z\right) +\left(
+\leftrightarrow \times \right) \right\}
\end{equation}

\begin{equation}
\frac{\delta h^{+}\left( \tau ,\vec y\right) }{\delta \phi \left( x^{\prime
}\right) }\mid _{\phi =0}=\frac{\left( H\tau ^{\prime }\right) ^2}{m_p}\int
d^3z\quad \Delta _{ret}\left( \left( \tau ,\vec y\right) ,\left( \tau
^{\prime },\vec z\right) \right) G_j^{+i}\left( \vec z-\vec x^{\prime
}\right) \partial _{ij}^2\varphi \left( x^{\prime }\right)
\end{equation}

The remaining steps consist of substituting these in (\ref{disker}),
computing the quantum expectation values with the help of the mode
decomposition  (\ref{modedecomp}). These straightforward manipulations
shall be omitted.

\subsection*{B. The Propagator Approach}

Let us recall the action for the microscopic $\varphi$ field

\begin{equation}
\label{sf2new}
S_f^{(2)}= \frac 12 \int d\tau d^3x ( H\tau)^{-2}
\left ( \varphi ^{\prime }-\vec \nabla \varphi ^2 \right ),
\end{equation}
where  the prime stands for a $\tau $ derivative.
We  derive from it the microscopic Feynman propagator \cite{Allen}.

\begin{equation}
\label{deltaf}\Delta _F(x_1,x_2)= -\left(\frac H{2\pi}\right) ^2
\left [ \frac{ \tau _1\tau _2 }{\left( \tau^2-r^2-i\epsilon \right) }+
 \frac 12 \log ( r^2-\tau ^2+i\epsilon ) -2c\right] ,
\end{equation}
where $  \tau  =\tau  _1-\tau _2$, $\vec r=\vec x_1-\vec x_2$, and
$c$ is an undetermined  constant.    The  retarded  propagator  $\Delta
_r=-2{\rm Im}\Delta _F\theta (\tau )$ and  the  Hadamard  propagator  $\Delta
_1=2{\rm Re}\Delta _F$ also follow.  .

The  graviton  Feynman  propagator  is  a  symmetric  bi-tensor
$\Delta_{j1,j2}^{i1,i2}(x_1,x_2)$. Out of symmetry considerations,
it must take the form  \cite{gravitonprop}

\begin{equation}
\label{deltagrav}{
\begin{array}{lll}
\Delta _{jl}^{ik} & = & {\Delta _{PP}}P_j^iP_l^k \\
& + & {\Delta _{PQ}}(P_j^iQ_l^k+Q_j^iP_l^k) \\
& + & {\tilde \Delta _{PQ}}(P^{ik}Q_{jl}+P_l^iQ^kj+Q^{ik}P_{jl}+Q_l^iP_j^k) \\
& + & {\Delta_{QQ}}Q_j^iQ_l^k \\
& + & {\tilde \Delta _{QQ}}(Q^{ik}Q_{jl}+Q_l^iQ_j^k)
\end{array}
}
\end{equation}
where

\begin{equation}
\label{plong}P^{ij}=\frac{\left( r^ir^j\right) }{r^2}
\end{equation}

\begin{equation}
\label{qtrans}Q^{ij}=\delta ^{ij}-P^{ij}
\end{equation}
The restrictions on physical gravitons imply a number of identities the
graviton propagator must satisfy, namely

\begin{equation}
\label{deltagauge}\Delta _{il}^{ik}=\Delta _{jl,i}^{ik}=0
\end{equation}
These identities allow us to write all the bi-scalar coefficients in terms
of $\Delta _{PP}$, concretely,

\begin{equation}
\label{deltacoef}{
\begin{array}{ccc}
{\Delta _{PQ}} & = & - \frac{1}{2} \Delta _{PP}\\
{\tilde \Delta _{PQ}} & = & \left( \frac {3}4+\frac{r^2}2\frac d{dr^2}\right)
\Delta _{PP} \\
{\Delta _{QQ}}& = & \left [ -\frac{9}8-3r^2\frac d{dr^2}-\left( r^2\frac
d{dr^2}
\right) ^2 \right ] \Delta _{PP} \\
{\tilde \Delta _{QQ}} & = & \left [ {\frac{11}8+3r^2\frac d{dr^2}+
\left( r^2\frac d{dr^2}\right) ^2} \right]
\Delta _{PP}
\end{array}
}
\end{equation}
Moreover, the  graviton  propagator  is  linked  to  the  scalar  one  by

\begin{equation}
\label{deltatr}\Delta _{ki}^{ik}=\left( \frac 8{m_p^2}\right) \Delta _F
\end{equation}
which provides a connection between $\Delta _{PP}$ and $\Delta _F$ , namely,

\begin{equation}
\label{deltapp}\Delta _{PP}= \frac{16}{3m_p^2} \int_0^1du\left[
u-u^4\right] \Delta _F\left( u\vec r\right)
\end{equation}

In  our approach here, the correlator  of  decohered
fluctuations is not to be guessed {\it a priori}, but should rather be
deduced from the analysis of the noise kernel induced by the inflaton-graviton
coupling. Our task  is simplified by the observation that this model
has the same structure as  the  $g\Phi  ^3$  theory
from  the  previous  section, only we now deal with a multicomponent field.
It  is  therefore  immediate  to  write down the noise kernel

\begin{equation}
\label{dsnoise}N(x_1,x_2)={\rm Re}\Sigma (x_1,x_2)
\end{equation}
and the dissipation kernel

\begin{equation}
\label{dsdiss}D(x_1,x_2)=2{\rm Im}\Sigma (x_1,x_2)\theta (\tau _1-\tau _2)
\end{equation}
where

\begin{equation}
\label{dssigma}\Sigma (x_1,x_2)=  \frac 2{( H\tau _1)^2 ( H\tau _2) ^2}
\left[ \partial _i\partial _j\partial
_k\partial _l \left( \Delta _F\Delta _{jl}^{ik} \right) \right] (x_1,x_2)
\end{equation}
An explicit evaluation yields
\begin{eqnarray}
\label{expsigma}
& & \Sigma (x_1,x_2)  =  {68\over (2\pi )^4m_p^2}{1\over \tau_1^4\tau_2^4}
{1\over (Z-1)^3} \Biggl \{ (Z-3) \Biggl [ (1-Z)^{-1}-\ln (\tau_1\tau_2(1-Z))-2c
\Biggr ] \nn \\
& & +   4b^2{(Z-5) \over (Z-1)^2} \Biggl [ {21 \over 100}-{a^2\over 15}
-{1\over 3b}- ( {a^4\over 5}+{a^2\over b} ) (1-a {\rm arctanh} (a^{-1})) \nn \\
& & + {3c\over 10} +({1\over 4}+{1\over 2b})\ln (1-{1\over a^2})-{3\over 20}
    \ln (\tau_1\tau_2(1-Z)) \Biggr ] \nn \\
& & + {2(Z-4)\over (Z-1)} \Biggl [(-b-2bc-(b+2)\ln (1-{1\over a^2})+
       b\ln (\tau_1\tau_2(1-Z)) \Biggr ]  \Biggl \}
\end{eqnarray}
where    $Z-1=\tau^2-r^2/2\tau_1\tau_2$,    $b=r^2/\tau_1\tau_2$,  and
$a^2=\tau^2/r^2$ (we assume $\tau^2$  has  a  small negative imaginary part to
obtain the correct time- ordering  property).    Observe  that  $Z$  is
de  Sitter invariant, while $a$ and $b$ are not. \\

\noindent {\bf Acknowledgement}
This work is supported in part by the National Science Foundation
under grants PHY91-19726, 
INT91-02088 (USA), and by CONICET, UBA and Fundaci\'on Antorchas (Argentina).
Thanks are due to N. Deruelle, Andrew Matacz and C. Stephens for discussions on
different aspects of this work.
BLH acknowledges support from the General Research Board of the
Graduate School of the University of Maryland and the Dyson Visiting Professor
Fund at the  Institute for Advanced Study, Princeton.
Part of this work was done while he visited the
Newton Institute for Mathematical Sciences at the University of Cambridge
during the Geometry and Gravity program in Spring 1994, and
the physics department of the Hong Kong University of Science and Technology
in Spring 1995.

\end{document}